\documentclass[prb,aps,showpacs,twocolumn]{revtex4}
\usepackage{mathrsfs}
\usepackage{pifont}
\usepackage{stmaryrd}
\usepackage{bbding}
\usepackage{amsmath}
\usepackage{amssymb}
\usepackage{latexsym}
\usepackage{graphicx}
\usepackage{dcolumn}
\usepackage{epstopdf}
\usepackage{amsmath,amssymb,epsf,bm}

\begin{document}
\title{Doping evolution of Zhang-Rice singlet spectral weight: a comprehensive examination by x-ray absorption spectroscopy}

\author{Y.-J. Chen,$^{1}$ M. G. Jiang,$^{2}$ C. W. Luo,$^{1}$ J.-Y. Lin,$^{2,\ast}$ K. H. Wu,$ ^{1,\dagger}$ J. M. Lee,$^{3}$ J. M. Chen,$^{3}$ Y. K. Kuo,$^{4}$ J. Y. Juang,$^{1}$  and Chung-Yu Mou$^{5,6}$}

\affiliation{$^1$Department of Electrophysics, National Chiao Tung University, Hsinchu 30010, Taiwan}
\affiliation{$^2$Institute of Physics, National Chiao Tung University, Hsinchu 30010, Taiwan}
\affiliation{$^3$National Synchrotron Radiation Research Center, Hsinchu 30076, Taiwan}
\affiliation{$^4$Department of Physics, National Dong Hwa University, Hualien 97401, Taiwan}
\affiliation{$^5$Department of Physics, National Tsing Hua University, Hsinchu 30043, Taiwan}
\affiliation{$^6$Physics Division, National Center for Theoretical Sciences, P.O. Box 2-131, Hsinchu, Taiwan}

\begin{abstract}
The total spectral weight \textit{S} of the emergent low-energy quasipaticles in high-temperature superconductors is explored by x-ray absorption spectroscopy. In order to examine the applicability of the Hubbard model, regimes that cover from zero doping to overdoping are investigated. In contrast to mean field theory, we found that \textit{S} deviates from linear dependence on the doping level \textit{p}. The slope of \textit{S} versus \textit{p} changes continuously throughout the whole doping range with no sign of saturation up to \textit{p} = 0.23. Therefore, the picture of Zhang-Rice singlet remains intact within the most prominent doping regimes of HTSC's.
\end{abstract}

\maketitle
\section{INTRODUCTION}

The high-temperature superconductors (HTSC's) have still attracted intensive research attention in condensed matter physics even after nearly three decades of their discovery. In some respects, HTSC's are even more elusive nowadays due to the very recent experimental endeavors and controversies. For examples, the issues such as the possible Fermi surface reconstruction \cite{D. LeBoeuf et al 2007} and a new competing order of charge ordering \cite{G. Ghiringhelli et al 2012} are reshaping our understanding of HTSC's. More fundamentally, the appropriate picture of doped quasiparticles in HTSC's and their orbital character have not completely emerged yet. Many believe that the central piece of these issues lies in the elucidation of the doped Mott insulator. The general consensus has been that the parent compound of HTSC's is a Mott insulator and the doped holes can be described by an effective Hubbard model or the \textit{t}-\textit{J} model projecting out the double occupancy. As holes are doped into the oxygen orbital in HTSC's, they hybridize with Cu spins to form the Zhang-Rice singlets (ZRS) \cite{F. C. Zhang et al 1988} residing between the upper Hubbard band (UHB) and the lower Hubbard band (LHB).

However, these foundations to our understanding of HTSC's have been challenged for the past several years.\cite{A. Comanac et al 2008} A noticeable case from the x-ray absorption spectroscopy (XAS) was proposed by Peets \textit{et al}.\cite{D. C. Peets et al 2009} By compiling available XAS data for different cuprates, the spectral weight of ZRS was claimed to be saturated for the hole doping level \textit{p} $\geq$ 0.21. (Also, see an earlier statement of saturated ZRS weight with \textit{p} $>$ 0.16 in Ref. 6.) This surprising finding of saturation in XAS explicitly indicated the breakdown of ZRS. Moreover, this early saturation, if held, actually indicated a failure of the three-band model of cuprates, as it was later pointed out.\cite{X. Wang et al 2010} Partly stimulated by this unexpected breakdown of the fundamental frame of Mott physics for HTSC's, very recently there have been many reports using modern theoretical techniques to reinvestigate this issue.\cite{X. Wang et al 2010,A. Liebsch 2010,P. Phillips et al 2010,T. Ahmed et al 2011,C.-C. Chen et al 2013} Despite of these new theoretical efforts, the conclusions drawn from these calculations are not entirely consistent with each other. Therefore, a new and more comprehensive experimental examination on Zhang-Rice singlet XAS is certainly indispensable.

In this work, we examine the XAS at the O \textit{K}- and Cu \textit{L}-edges using well (001)-oriented $\rm Y_{1-\textit{x}}Ca_{\textit{x}}Ba_{2}Cu_{3}O_{7-\delta}$ thin films within the 123 family to avoid any possible discrepancy due to different cuprate systems. By analyzing the total spectral weight revealed in the Zhang-Rice band of the O \textit{K}-edge and ligand-holes in the Cu \textit{L}-edge, we show that, in consistent with theoretical calculations, both spectral weights show clear deviations from the simple linear dependence on the doping level \textit{p}. In addition, we find that the slope of \textit{S} versus \textit{p} changes continuously throughout the whole doping range without any sign of saturation up to \textit{p} = 0.23. \textit{Our results render the validity of Zhang-Rice singlet within the most intensively studied doping regimes of HTSC's}. 

\section{EXPERIMENTS}
We start by characterizing samples and measurements performed in this work. Highly oriented (001) $\rm Y_{1-\textit{x}}Ca_{\textit{x}}Ba_{2}Cu_{3}O_{7-\delta}$ (Ca-YBCO; \textit{x} = 0, 0.3 and 0.4) thin films were deposited on (100) $\rm SrTiO_{3}$ substrates by the pulsed laser deposition (PLD) method. The KrF excimer laser ($\lambda$ = 248 nm) operating at a repetition rate of 5 Hz with an energy density of $\sim$5 $\rm J/cm^{2}$ was used for the Ca-YBCO film growth in the optimal growth condition of 740$-$770 $\rm ^\circ C$ substrate temperatures and 0.26$-$0.3 Torr oxygen partial pressures. The electrical resistivity as a function of temperature $\rho(\it T)$ was measured by a standard four-probe method. The oxygen contents of the films can be repeatedly changed by controlling temperatures and oxygen pressures inside the quartz tube surrounded by the furnace.\cite{K. H. Wu et al 1998} The XAS at the O \textit{K}-  and Cu \textit{L}-edges was performed in fluorescence mode using synchrotron radiation from a 6 m high-energy spherical monochromator (HSGM) beam line at National Synchrotron Radiation Research Center in Taiwan. Details of XAS experiments can be found elsewhere.\cite{S. J. Liu et al 2003,C. W. Luo et al 2003} All thin films were about 300 nm thick to avoid the absorption contribution from the oxide substrates. The self-absorption corrections were applied to these spectra to correct the saturation effects. The energy resolution of the monochromator was set to $\sim$0.1 eV for the O \textit{K}-edge range and $\sim$0.2 eV for Cu \textit{L}-edge.

The doping level \textit{p} was determined with special care. For superconducting samples, the estimated value of \textit{p} was derived from $1 - {{{T_c}} \mathord{\left/
 {\vphantom {{{T_c}} {{T_{c,\max }} = 82.6{{(p - 0.16)}^2}}}} \right.
 \kern-\nulldelimiterspace} {{T_{c,\max }} = 82.6{{(p - 0.16)}^2}}}$,\cite{J. L. Tallon et al 1995} where ${T_{c,\max }}$ is the maximum transition temperature at the optimum doping in each system. For heavily underdoped thin films of \textit{p} = 0.02$-$0.05 with no $T_{c}$, \textit{p} was determined from the value of the thermoelectric power (TEP) at room temperature.\cite{J. L. Tallon et al 1995} The \textit{p} values of several superconducting underdoped or overdoped samples were cross checked by both $T_{c}$ and the thermoelectric power. The results of \textit{p} were found to be consistent between these two methods. For TEP measurements, the magnitude of the temperature gradient was maintained below 1 K and the sample space was maintained in a good vacuum (10$^{-4}$ Torr). Due to the high resistivity in the extremely underdoped regime of Ca-YBCO, the \textit{p} value of the second lowest doping sample was assessed to be about 0.012 from the interpolation of the peak intensity of UHB in O \textit{K}-edge XAS.

\section{RESULTS AND DISCUSSION}
\begin{figure}
\begin{center}
\includegraphics[width=8.5cm]{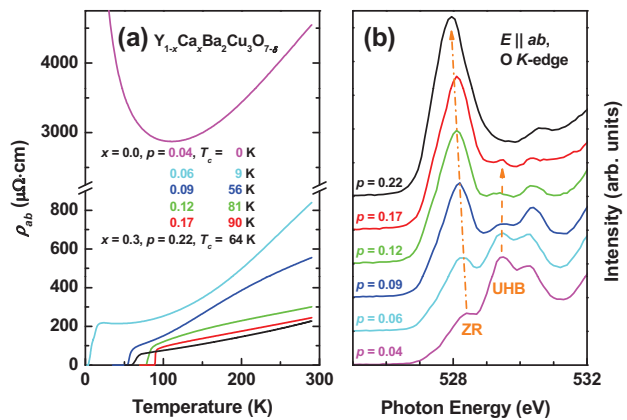}
\caption{\label{fig:FIG1}(color online) (a) The temperature dependence of in-plane resistivity (${\rho_{ab}}$) for YBCO (\textit{x} = 0) thin films with various hole concentrations (\textit{p} = 0.04, 0.06, 0.09, 0.12 and 0.17) and Ca-YBCO (\textit{x} = 0.3) thin film for \textit{p} = 0.22. (b) The polarized (\textit{E}$\parallel$\textit{ab}) XAS of O \textit{K}-edge of YBCO and Ca-YBCO thin films measured at room temperature. The yellow dashed and dash-dot lines are guide to eyes to depict that the ZRS and UHB are moving away from each other in energy as \textit{p} increases.}
\end{center}
\end{figure}

\begin{figure}
\begin{center}
\includegraphics*[width=8.5cm]{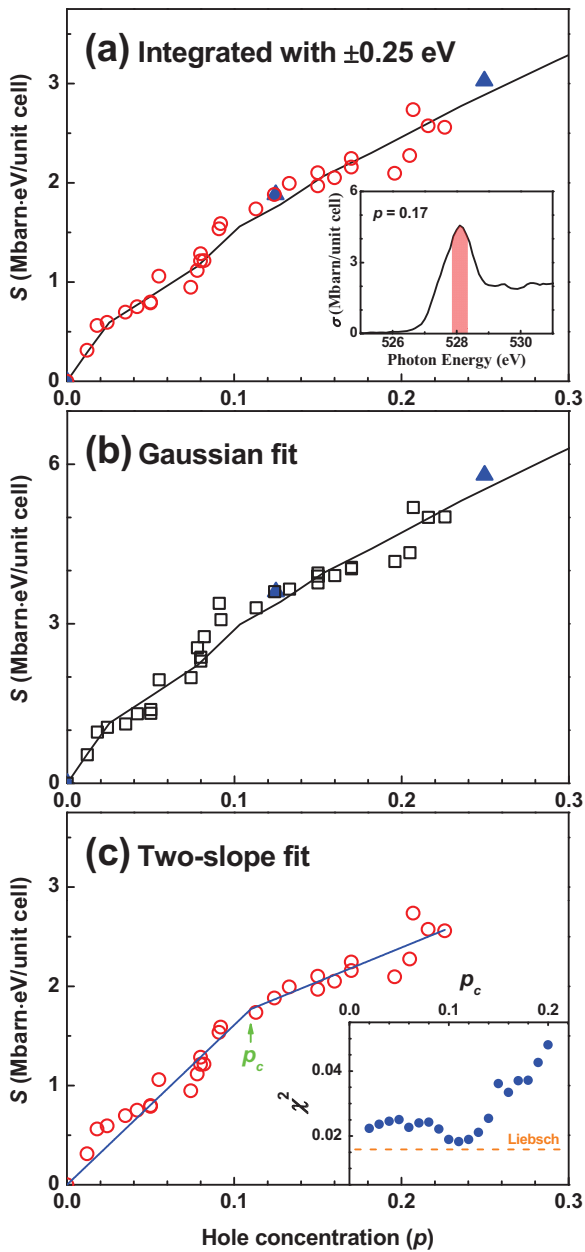}
\caption{\label{fig:FIG2}(color online) (a) The doping \textit{p} dependence of the spectral weight \textit{S} of ZRS (red open circles) from the integration within $\pm$0.25 eV of the ZRS peak energy. The red area shown in the inset is an example integration for \textit{p} = 0.17 YBCO. The black curve was calculated by Liebsch.\cite{A. Liebsch 2010} Theoretical calculations (blue triangles) from Chen \textit{et al}. \cite{C.-C. Chen et al 2013} normalized at \textit{p} $\sim$0.125 were also depicted. (b) $S(\it p)$ (black open squares) from the multi-peak Gaussian fit. (c) $S(\it p)$ analyzed by the two-slope fit. The inset shows the chi-squared distribution (${\chi ^2}$) with the turning point ($p_{c}$) as well with the ${\chi ^2}$ value from the curve fit of Liebsch (the orange dashed line).}
\end{center}
\end{figure}

Fig. 1(a) shows $\rho(\it T)$ of the representative samples from underdoping to overdoping. The \textit{T} dependence and the values of $\rho(\it T)$ both demonstrate the high quality of the thin films nearly as good as that of single crystals. Fig. 1(b) shows the corresponding O \textit{K}-edge XAS spectra of YBCO and Ca-YBCO samples in Fig. 1(a). XAS provides information of the unoccupied states near the Fermi energy $E_{F}$, and is considered as one of the most appropriate tools to explore the doped Mott insulators. Actually, early seminal XAS work had helped establish the ZRS picture in HTSC's.\cite{C. T. Chen et al 1992,C. T. Chen et al 1991} The features of the O \textit{K}-edge XAS around \textit{E} = 527.5 eV, 528.2 eV, and 529.4 eV are attributed to the CuO chain holes, ZRS, and UHB, respectively. The ZR band and the UHB are not rigid and are correlated to each other. Removing \textit{p} electrons from a Mott insulator in the atomic limit should increase the ZRS spectral weight by 2\textit{p} at the expense of UHB spectral weight by \textit{p}, as qualitatively shown in Fig. 1(b). (LHB also loses the spectral weight by \textit{p}, which can not be detected by XAS.) The peaks of ZRS and UHB evolve away from each other with increasing \textit{p} as suggested by some calculations that the UHB shifts away from $E_{F}$ when \textit{p} becomes larger.\cite{X. Wang et al 2010,A. Liebsch 2010,C.-C. Chen et al 2013}

The main theme of the present work is the \textit{p} dependence of the ZRS spectral weight \textit{S}. Fig. 2(a) shows $S(\it p)$ from \textit{p} = 0 to \textit{p} = 0.23, covering both the underdoped and overdoped regimes. The open red circles denote the experimental spectral weight from the integration within $\pm$0.25 eV of the ZRS peak energy as illustrated by the inset of Fig. 2(a). (The data point at \textit{p} = 0 came from the zero ZRS spectral weight of a sample rather than the extrapolation.) The earlier calculations have pointed out that the spectral weight \textit{S} should be linear with respect to the doping level \textit{p} in the present doping range.\cite{H. Eskes et al 1990,M. B. J. Menders et al 1993} However, a recent experimental XAS study claimed a saturation of \textit{S} for \textit{p} $\geq$ 0.21.\cite{D. C. Peets et al 2009} As seen in Fig. 2(a), our results do show a weak slope change in \textit{S} throughout the whole doping range, but no sign of saturation was observed. The recent calculations via the two-dimensional one-band Hubbard model (black solid line) \cite{A. Liebsch 2010} and three-orbital Hubbard model (blue solid triangles) \cite{C.-C. Chen et al 2013} were also depicted in Fig.2 (a). It is seen that both one-band and three-orbital Hubbard models capture the trend of $S(\it p)$ with impressive accuracy. According to Ref. 11, this slope change is due to the combined effects of core-hole interaction and weakened correlation with doping. The doping levels from \textit{p} = 0 to \textit{p} = 0.23 are the regimes that host superconductivity and other competing orderings like the pseudogap for HTSC's. Therefore, the Hubbard model and ZRS seem to be good (at least to the first order) approximations to describe the intriguing natures of HTSC's. For the present studied doping regimes, the one-band and the three-orbital Hubbard models lead to almost the same results (see Fig. 2(a)). In principle, the three-orbital Hubbard model would go beyond the one-band Hubbard model and predict a smaller \textit{S} at higher doping than the one-band model does.\cite{C.-C. Chen et al 2013} In principle, an effective doping dependent Hubbard $U(\it p)$ is needed to fit the spectra weight.\cite{C.-C. Chen et al 2013,R. S. Markiewicz et al 2010} Further comprehensive examinations at higher doping levels are certainly desirable.

Moreover, in the low doping regimes \textit{S} also shows features rather than a simple linear behavior respect to the doping level \textit{p} (see Fig. 2(a)). Explicitly, there exists an unexpected kink at \textit{p} $\approx$ 0.02 in $S(\it p)$. The slope of $S(\it p)$ suddenly changes into a smaller value for \textit{p} $>$ 0.02. Experimentally, this kink might manifest the interplay between O 2\textit{p} and Cu 3\textit{d} states as shall be discussed later. Interestingly, the one-band model appears to capture the kink feature as indicated by the black solid line in Fig. 2(a).\cite{A. Liebsch 2010} More theoretical studies are needed to fully identify the origin of the surprising slope change at such a low doping level.

\begin{figure}
\begin{center}
\includegraphics*[width=8.5cm]{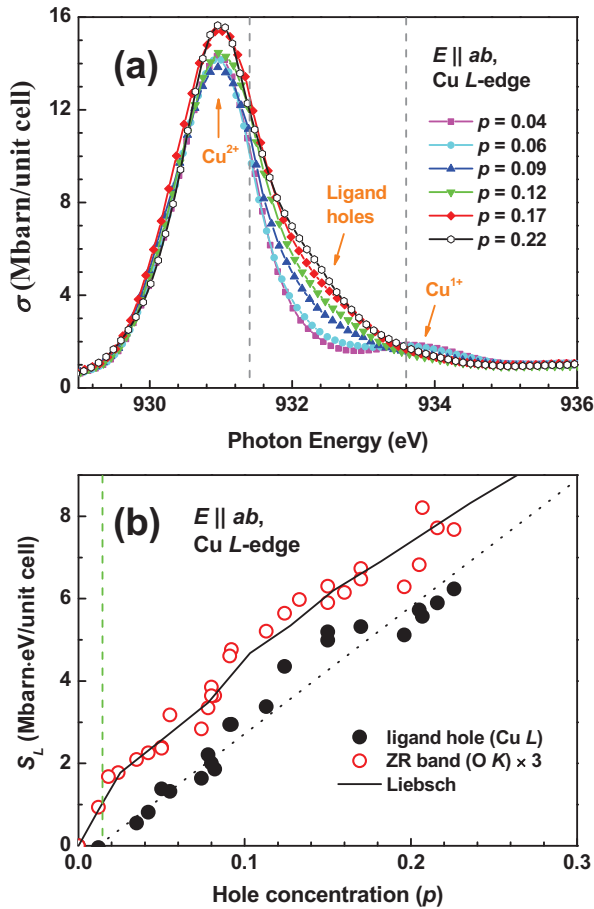}
\caption{\label{fig:FIG3}(color online) (a) The polarized (\textit{E}$\parallel$\textit{ab}) Cu \textit{L}-edge XAS of YBCO and Ca-YBCO thin films measured at room temperature. The spectra consist of a main peak near 931 eV (associated with Cu$^{2+}$), a broad peak near 932 eV (ligand holes), and a small peak near 934 eV (related to Cu$^{1+}$ on empty Cu-O chains). (b) The spectra weight $S_{L}$ of the ligand hole as a function of \textit{p} was estimated by integrating the area within two gray dashed lines (as shown in Fig. 3a). The black dot line is the linear fit of $S_{L}(\it p)$. The green dashed line indicates the doping level below which $S_{L}(\it p)$ is negligible.}
\end{center}
\end{figure}

There has been concern that different integration energy windows might lead to different conclusions.\cite{D. C. Peets et al 2009,X. Wang et al 2010,A. Liebsch 2010,P. Phillips et al 2010,T. Ahmed et al 2011,C.-C. Chen et al 2013} We also tried a $\pm$0.5 eV integration window and basically obtained the same \textit{p}-dependence of $S(\it p)$ (results not shown). Furthermore, as shown in Fig. 2(b), $S(\it p)$ of ZRS band was obtained from the Gaussian multi-peak fit. Experimentally, the integration window should not a major factor since the spectra inevitably have suffered instrument broadening. $S(\it p)$ from the Gaussian peak fit is nearly identical to that of from the energy window integrating as seen in Fig. 2(a) and (b). To avoid being biased by existing theoretical calculations, a phenomenological two-slope fit was further used to analyze the data as shown in Fig. 2(c). In this two-slope fit, $p_{c}$ is the doping level at which the slope changes. The blue solid line in Fig. 2(c) represents the optimal two-slope analysis with $p_{c}$ = 0.11. While the fitting quality of the two-slope analysis is inferior to that of Liebsch's curve as shown in Fig. 2(c) inset, in either case it suggests that the slope of $S(\it p)$ changes within the present doping regimes.

Fig. 3(a) shows the corresponding Cu \textit{L}-edge XAS spectra of YBCO and Ca-YBCO samples in Fig. 1(a). It is noted that the Cu \textit{L}-edge XAS with selective \textit{p} samples are shown here. In addition to the Cu$^{2+}$ and Cu$^{1+}$ features, in between them there exists an extra spectral weight. This feature has also been considered to be relevant to ZRS and attributed to the ligand hole states in CuO$_{2}$ planes.\cite{C. T. Chen et al 1992,M. Merz et al 1998} The spectral weight $S_{L}(\it p)$ was extracted by integration of ${\sigma _{\rm Cu}}(p) - {\sigma _{\rm Cu}}(p = 0)$ between \textit{E} = 931.4 eV and 933.6 eV as denoted by two vertical dashed lines in Fig. 3(a). $S_{L}(\it p)$ was further corrected to the Gaussian fit of the Cu$^{1+}$ feature and plotted in Fig. 3(b). The subsequent $S_{L}(\it p)$ was plotted together with $S(\it p)$ of O \textit{K}-edge in Fig. 2(a) for comparison. At \textit{p} = 0.012, ZRS has possessed significant spectral weight, while $S_{L}$ is still near zero. This indicates that the doped holes exclusively enter O 2\textit{p} states at a very small doping. It is noticed that spectral weight $S_{L}(\it p)$ shows nonzero value when the doping level \textit{p} $>$ 0.02, presumably due to the fact that doped holes starts entering Cu 3\textit{d} states. Consequently, the N$\rm \acute{e}$el temperature $T_{N}$ significantly decreases to zero for \textit{p} $>$ 0.02 since the long range antiferromagnetic order is frustrated by the neighboring unoccupied sites. The evolution of the orbital character of the doped holes with \textit{p} has been recently demonstrated by Compton scattering.\cite{Y. Sakurai et al 2011} Therefore, the experimental reason for the kink at \textit{p} $\approx$ 0.02 and the slope change of $S(\it p)$ for the entire doping regime could be that the O 2\textit{p} states lose part of its spectral weight to Cu 3\textit{d} states due to the stronger Cu 3\textit{d} character of doped holes with increasing \textit{p}. Comparing the linear fit of the $S_{L}$ data (black dot line) with the solid line in Fig. 3(b), $S_{L}(\it p)$ has a weaker slope change than $S(\it p)$ does. This observation is also consistent with the stronger Cu character of the doped holes with increasing \textit{p} in the YBCO system.

\section{CONCLUSIONS}

To conclude, the Hubbard model and Zhang-Rice singlets describe the spetral weight of doped quasiparticles $S(\it p)$  well in high-temperature superconductors for a wide doping range. This doping regime, as presented in this study, covers almost all interesting states in high-temperature superconductors including antiferromagnetism, superconductivity, pseudogap, stripe phase, and charge ordering. To distinguish the validity of three-orbital Hubbard model from  that of the one-band model, x-ray absorption spectroscopy measurements with even higher doping levels are needed.

\begin{acknowledgements}
We thank D.-J. Huang and T. K. Lee for useful discussions. This work was supported by the National Science Council of Taiwan, R.O.C. under grants: NSC-101-2112-M-009-017-MY2, NSC-101-2112-M-009-020, NSC-98-2112-M-009-016-MY2, NSC-100-2112-M-259-002-MY3, and MOE ATU program at NCTU.\\
\end{acknowledgements}

\email [{$^{\ast}$ago@cc.nctu.edu.tw}

\email [{$^{\dagger}$khwu@cc.nctu.edu.tw}

\end{document}